\documentstyle[prd,aps,floats]{revtex}

\begin{document}
\preprint{SUSSEX-AST 96/7-7, astro-ph/9607166}
\draft

%
%
\input epsf
\renewcommand{\topfraction}{0.8}
\twocolumn[\hsize\textwidth\columnwidth\hsize\csname 
@twocolumnfalse\endcsname

\title{Open inflationary universes in the induced gravity theory}
\author{Anne M. Green and Andrew R. Liddle}
\address{Astronomy Centre, University of Sussex, Falmer, Brighton BN1 
9QH,~~~U.~K.}
\date{\today}
\maketitle
\begin{abstract}
The induced gravity theory is a variant of Jordan--Brans--Dicke theory where 
the `dilaton' field possesses a potential. It has the unusual feature that 
in the presence of a false vacuum there is a {\em stable} static solution 
with the dilaton field displaced from the minimum of its potential, giving 
perfect de Sitter expansion. We demonstrate how this solution can be used to 
implement the open inflationary universe scenario. The necessary second 
phase of inflation after false vacuum decay by bubble nucleation is driven 
by the dilaton rolling from the static point to the minimum of its 
potential. Because the static solution is stable whilst the false vacuum 
persists, the required evolution occurs for a wide range of initial 
conditions. As the exterior of the bubble is perfect de Sitter space, there 
is no problem with fields rolling outside the bubble, as in one of the 
related 
models considered by Linde and Mezhlumian, and the expansion rates before 
and after tunnelling may be similar which prevents problematic 
high-amplitude super-curvature modes from being generated. Once 
normalized to the microwave background anisotropies seen by the COBE 
satellite, the viable models form a one-parameter family for each 
possible $\Omega_0$.
\end{abstract}
\pacs{PACS numbers: 98.80.Cq \hspace*{2cm} Sussex preprint SUSSEX-AST 
96/7-7, astro-ph/9607166}

\vskip2pc]

\section{Introduction}

Models of inflation leading to an open universe, which have a surprisingly 
long history \cite{gott,gs,gott2}, have received a new lease of life 
recently \cite{styy,bt,lin}. The idea relies on the observation that the 
inside of a bubble formed by quantum tunnelling looks exactly like an open 
universe to an observer inside the bubble \cite{CdL}, plus the realization 
that if a period of inflation occurs within the bubble after tunnelling it 
is possible to tune the present value of the density parameter inside the 
bubble to whatever value one desires, for example $\Omega_0 \simeq 0.3$ as 
favoured by certain types of cosmological observation. An initial phase of 
inflation before tunnelling is also required, in order to provide a 
homogeneous background within which the bubble nucleation can occur.

The earliest specific realizations of the open inflation idea utilized a 
single scalar field with a rather complicated effective potential 
\cite{gott2,bt}. The potential requires a local minimum at non-zero energy, 
in which the trapped scalar field drives the first phase of inflation, plus 
a flat region of the potential on the other side of the barrier, down which 
the field can slowly roll after tunnelling to drive the second 
inflationary period. This second period must have a duration around $60$ 
$e$-foldings; much more and the universe will end up flat, much less and it 
will end up empty. While implementing the open inflation paradigm with just 
a single field is in principle an attractive possibility, the perceived 
unnaturalness of a potential with so many features mitigates against it.

A much more natural framework for open inflation was introduced in 
Ref.~\cite{lin}, where two scalar fields were used. In this case, the field 
which drives the second period of inflation is distinct from the field 
undergoing the tunnelling. The two fields need not even be explicitly 
coupled --- the potential
\begin{equation}
\label{linpot}
V(\phi,\sigma) = V(\sigma) + \frac{1}{2} m^2 \phi^2
\end{equation}
can be used \cite{lin}, where $V(\sigma)$ is a potential featuring a 
metastable vacuum. The $\phi$ field rolls down the potential from large 
$|\phi|$ while $\sigma$ is in the false vacuum driving the first period of 
inflation, then at some point $\sigma$ tunnels generating the open universe, 
and then $\phi$ continues to roll to the bottom of its potential driving the 
second phase of inflation as it goes. Linde and Mezhlumian \cite{lin} 
discuss several variants of this basic scheme.

Introducing the second field, while improving the naturalness of the models,
can bring in problems of its own. The main one is that the space-time {\em 
outside} the bubble may no longer be perfectly de Sitter, due to evolution 
of the non-tunnelling field. This breaks the homogeneity at the bubble wall  
which forms the initial ($t=0$) hypersurface for the open universe within 
the bubble \cite{lin}, even in the absence of bubble-wall fluctuations which 
we discuss later. In the model quoted above, this is a serious problem 
unless $\phi$ can be made to roll much more slowly outside the bubble than 
inside. This can be achieved by ensuring that the energy density after 
tunnelling is much less than that before. However, even this solution may 
bring a new problem, discussed by Linde and Mezhlumian \cite{lin} 
and computed in detail by Sasaki and Tanaka \cite{st}, that long-wavelength 
modes, known as super-curvature modes, may be excessively generated if the 
difference in energy density before and after tunnelling is large. This 
problem is avoided in two other models in Ref.~\cite{lin}, where the fields 
are static before tunnelling.

In this paper we describe a new model for open inflation, based on the 
induced gravity theory \cite{z}. In this theory, which is basically the 
Jordan--Brans--Dicke theory \cite{bd,will} with a symmetry-breaking 
potential for the Brans--Dicke, or dilaton, field $\phi$, the field driving 
the second period of inflation is associated with the gravitational sector 
of the theory rather than introduced by hand into the matter sector. The 
tunnelling field, as in the models above, is to be associated with some 
symmetry breaking in the early universe.

This model is ideally suited for the implementation of open inflation, 
because in the presence of a false vacuum there is a stable, static solution 
for the dilaton where it is displaced from the minimum of its potential. All 
initial conditions evolve into this solution, which acts as a late-time 
attractor, as long as the false vacuum persists. It corresponds to a perfect 
de Sitter solution, and thus is an ideal environment for nucleation to take 
place. When the false vacuum decays, the dilaton suddenly `realizes' that it 
is displaced from the minimum of its potential and slow-rolls into it, 
driving the necessary second period of inflation.

As we shall see, this scenario can be implemented for very reasonable 
choices of parameters. Once one normalizes the density perturbations 
produced by quantum fluctuations to the observations by the COBE satellite 
\cite{COBE}, one is left with a one-parameter family of possible models for 
each value of the present density parameter $\Omega_0$. Because there is no 
problem of rolling of the fields outside the bubble, there is no necessity 
for a large energy difference before and after tunnelling, and so no danger 
of excessive super-curvature modes.

\section{The Model and the Static Solution}

The action for induced gravity is \cite{z}
\begin{equation}
S = \int {\rm d}^4 x \sqrt{-g} \left[ \frac{1}{2} \xi \phi^2 R 
	- \frac{1}{2} \partial_{\mu} \phi \partial^{\mu} \phi  
	+ V(\phi) + {\cal L}_{{\rm mat}} \right] \,,
\end{equation}
where we shall refer to the field $\phi$ as the dilaton. In our conventions, 
an Einstein--Hilbert term requires a positive sign, so $\xi$ must be 
positive and the effective gravitational coupling is $G_{{\rm eff}}=1/(8 \pi 
\xi \phi^2)$. The choice of potential for $\phi$ is not very important for 
our purposes; we choose the simplest symmetry breaking form \cite{z}
\begin{equation}
V(\phi) = \frac{1}{8} \lambda ( \phi^2 - \nu^2 )^2 \,.
\end{equation}
In order that the correct present strength of gravity is obtained when the 
$\phi$ field sits in its minimum, we require
\begin{equation}
\label{nu}
\nu^2 = m_{{\rm Pl}}^2/8\pi \xi \,.
\end{equation}
This action is in fact exactly that of Jordan--Brans--Dicke theory 
\cite{bd,will} with a potential for the Brans--Dicke field; the 
correspondence is given by, in the usual notation
\begin{equation}
\Phi=8 \pi \phi^2 \quad , \quad \omega=\frac{1}{4\xi} \,.
\end{equation}
Because of the presence of the potential, the usual solar system and 
nucleosynthesis limits on $\omega$ do not apply.

The matter lagrangian ${\cal L}_{{\rm mat}}$ contains all the other matter 
in the theory. We shall assume that during the initial inflationary phase 
this is dominated by a single scalar field $\sigma$
\begin{equation}
{\cal L}_{{\rm mat}} = -\frac{1}{2} \partial_{\mu} \sigma
	\partial^{\mu} \sigma + V(\sigma) \,,
\end{equation}
whose potential possesses both a metastable vacuum state and a true minimum. 
The detailed form of the potential is irrelevant; all we need to know is the 
energy of the false vacuum, $V_{{\rm fv}}$, and the tunnelling rate $\Gamma$ 
from this state to the true vacuum.

Previously, inflation during induced gravity symmetry breaking has mainly 
been considered with only one scalar field, the dilaton, and hence a single 
period of inflation \cite{azt,Kaiser,fu}. Accetta and Trester \cite{at}, 
however, have considered the action above in the context of extended 
inflation, the essential difference being that they require percolation of 
the true vacuum bubbles whereas we require that the bubble nucleation rate 
is sufficiently small that the bubbles remain isolated.   

The Friedmann equation and the equation of motion for $\phi$ are
\begin{eqnarray}
& & H^2 + 2 H \frac{\dot{\phi}}{\phi} + \frac{k}{a^2} = 
	\frac{1}{3 \xi \phi^2} \left[ \frac{ \dot{\phi}^2}{2} + V(\phi)
	+ V(\sigma) \right] \,, \\
\label{peqn}
& & \ddot{\phi} + 3H \dot{\phi} + \frac{ \dot{\phi}^2}{\phi} =  
	\frac{1}{1 + 6 \xi} \left[ \frac{4 ( V(\phi) + 
	V(\sigma))}{\phi} - \frac{ {\rm d} V}{{\rm d} \phi} \right] \,.
\end{eqnarray}

Remarkably, while the $\sigma$ field is in the false vacuum these equations 
possess an exact {\em static} solution for a particular value of $\phi^2$, 
driving a de Sitter expansion. To our knowledge, this solution has not 
appeared before in the literature.\footnote{Accetta and Trester \cite{at} 
considered exactly this action in a different context and displayed a 
solution of this type, but they appear to have derived it incorrectly (their 
Eq.~(5) is missing a term).} This occurs when the right hand side of 
Eq.~(\ref{peqn}) equals zero; labelling the appropriate value of $\phi$ as 
$\phi_{{\rm st}}$, it is given by
\begin{equation}
\label{phist}
\phi_{{\rm st}}^2 = \nu^2 \left( 1 + \frac{8V_{{\rm fv}}}{\lambda \nu^4}
	\right) \,,
\end{equation}
leading to (at late times once the curvature term becomes negligible) the 
expansion rate
\begin{equation}
\label{Hst}
H_{{\rm st}}^2 = \frac{8\pi V_{{\rm fv}}}{ 3 m_{{\rm Pl}}^2} \,. 
\end{equation}
Interestingly, this looks exactly like the usual Friedmann equation, but 
note that it is the {\em present} Planck mass which appears in this formula 
and not the effective Planck mass at that time, given by $\sqrt{8\pi \xi 
\phi_{{\rm st}}^2}$. Note also that this solution only exists when the false 
vacuum is present; if $V_{{\rm fv}} = 0$ then the only static solution is 
$\phi_{{\rm st}} = \nu$ corresponding to a completely empty 
universe.\footnote{However, other contributions to ${\cal L}_{{\rm mat}}$, 
such as non-rela\-tiv\-istic matter, can also displace $\phi$ from its 
minimum at later epochs.}

\begin{figure}[t]
\centering 
\leavevmode\epsfysize=6cm \epsfbox{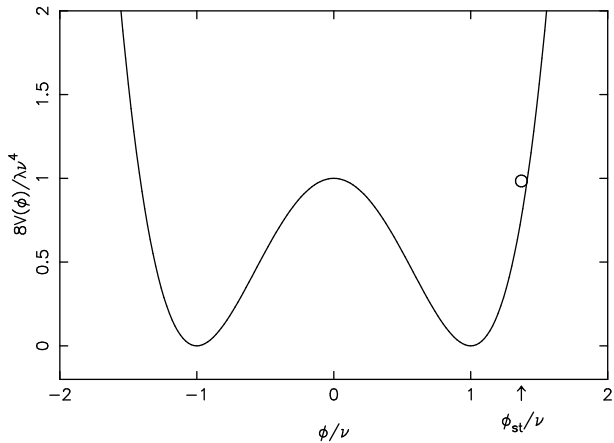}\\ 
\caption[fig1]{\label{fig1} The potential 
for the dilaton, with minima at $\phi^2 = \nu^2$. The location of the static 
point $\phi_{{\rm st}}$, here taken to be positive and indicated by the 
circle, depends on the size of the false vacuum. When the false vacuum 
decays, $\phi$ rolls to its true minimum, driving a second period of 
inflation.} 
\end{figure} 

An important feature of this solution is that it is stable. This is 
intuitively obvious (there being no other type of solution into which it can 
decay); we have also checked this explicitly by considering a linear 
perturbation about $\phi_{{\rm st}}$. The static solution is 
stable regardless of parameters, but the nature of the stability can 
change; defining $V_{{\rm fv}}^{{\rm crit}} = 4 \xi \lambda 
\nu^4/3(1+6\xi)$, then for $V_{{\rm fv}} < V_{{\rm fv}}^{{\rm crit}}$ the 
perturbation oscillates with exponentially decaying amplitude, while for 
$V_{{\rm fv}} > V_{{\rm fv}}^{{\rm crit}}$ it just decays exponentially. An 
alternative view of the stability is given below.

In fact, as long as the false vacuum persists the static solution is a 
late-time attractor for all initial conditions. For instance, even if $\phi$ 
is placed in its minimum at $\phi = \nu$, it will then drift up the 
potential until it reaches the static solution. Therefore all regions of 
space in which $\sigma$ is trapped for sufficiently long in its false vacuum 
reach the same physical state at late times. Fig.~1 illustrates this 
behaviour schematically.

The static solution displaced from the minimum of the potential seems 
unusual in the theory as written, but this is simply because there are extra 
terms, acting similarly to an effective potential, coming from the coupling 
to gravity, so that $V(\phi)$ is insufficient to determine the dynamics of 
the dilaton. An alternative viewpoint of the static solution comes from 
considering instead the conformally related Einstein frame, where the 
coupling of the dilaton to gravity is removed by the appropriate conformal 
transformation. We make the transformation 
\begin{equation}
\label{conftran}
{\rm d}t = C^{-1} {\rm d} \tilde{t} \quad , \quad a(t) = C^{-1}
	\tilde{a}(\tilde{t}) \,,
\end{equation}
where the required conformal factor is $C = \phi/\nu$. In order to have 
a canonical kinetic term, we can redefine the dilaton according to 
\begin{equation}
\label{fielddef}
\frac{\tilde{\phi}}{m_{{\rm Pl}}} = \sqrt{\frac{1 + 6 \xi}{32 \pi \xi}} 
	\; \ln \left( \frac{\phi^2}{\nu^2} \right) \,.
\end{equation}

\begin{figure}[t]
\centering 
\leavevmode\epsfysize=6cm \epsfbox{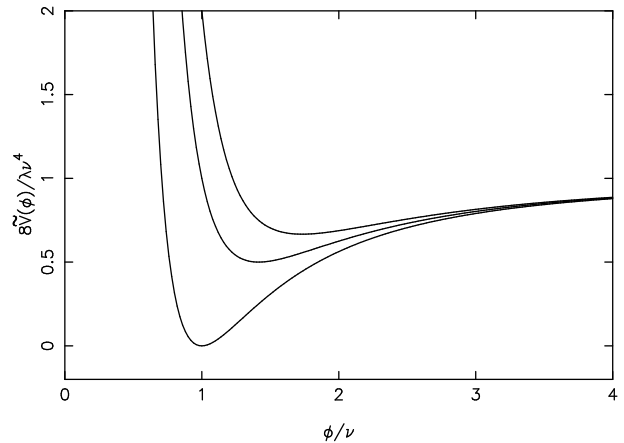}\\ 
\caption[fig2]{\label{fig2} The potential $\tilde{V}(\phi)$ of the dilaton 
in the Einstein frame, shown in normalized units for different choices of 
the false vacuum energy --- from bottom to top $8V_{{\rm fv}}/\lambda\nu^4 = 
0$, $1$ and $2$. The static point is at the minimum of $\tilde{V}(\phi)$, 
and is displaced to the right of $\phi = \nu$ by the presence of the false 
vacuum.} 
\end{figure}

In the Einstein frame, the potential for the dilaton is given by
\begin{equation}
\frac{\tilde{V}(\phi)}{\nu^4} = \frac{\lambda \left( \phi^2 - \nu^2
	\right)^2}{8\phi^4} + \frac{V_{{\rm fv}}}{\phi^4} \,,
\end{equation}
where we have left things in terms of the original $\phi$ field. The minimum 
of this potential is indeed at $\phi_{{\rm st}}$, and this is the only local 
minimum of the potential and hence the late-time attractor for all initial 
conditions, provided the false vacuum persists. The nature of the stability 
as given above can also be derived in this framework. Fig.~2 shows 
$\tilde{V}(\phi)$ for a selection of false vacuum energies. 

\section{Inflation after tunnelling}

\subsection{Creating an open universe}

The static solution provides the initial environment in which open inflation 
can be realized, by providing a large homogeneous region. An open universe 
is now created by tunnelling of the $\sigma$ field to its true vacuum, 
leaving only the gravitational sector of the theory. With $V_{{\rm fv}}$ now 
gone, the static point loses its support and the $\phi$ field `notices' that 
it is displaced from its true minimum. It therefore begins to evolve towards 
it. Because $V(\phi)$ is quite a flat potential, inflation can proceed while 
it does so. The interesting situation is where just enough inflation occurs 
inside the bubble so as to generate an open, but not too empty, universe at 
the present. The precise amount of inflation required to do this depends 
somewhat on the energy scale of inflation and on the details of reheating; 
for definiteness we shall simply assume that $60$ $e$-foldings of inflation 
in the second epoch are sufficient to place the present density parameter 
$\Omega_0$ somewhere in the interesting range $0.1 < \Omega_0 < 0.9$ 
\cite{bt}.

Inflation in the induced gravity theory without a false vacuum, now 
essentially just a single field model, has already been investigated 
\cite{azt,Kaiser,fu}. We simply need to locate the point on the potential 
corresponding to 60 $e$-foldings from the end, $\phi_{60}$, and ensure that 
it corresponds to the initial condition for the second epoch. That is, we 
arrange that $\phi_{{\rm st}} = \phi_{60}$, by making the appropriate 
choice of the false vacuum energy $V_{{\rm fv}}$ in Eq.~(\ref{phist}). A 
displacement of $V_{{\rm fv}}$ from this criterion alters the initial 
condition for the second phase, and hence results in a different $\Omega_0$. 
Any value of the present density parameter can therefore be arranged by 
making the appropriate choice of false vacuum energy.

Notice that in this model, unlike those discussed in Ref.~\cite{lin}, every 
bubble which nucleates gives rise to the same value of $\Omega_0$, assuming 
that the parameters of the underlying theory are fixed and that the region 
of the universe nucleated from the static solution. In Ref.~\cite{lin}, 
models were discussed in which different bubbles could have different 
density parameters, giving an ensemble of different density universes.
 
\subsection{Avoiding bubble percolation}

In order to obtain a homogeneous open universe we require that the bubbles 
of true vacuum produced are isolated; we don't want our bubble universe to 
have already collided with others. The nucleation rate per Hubble volume per 
Hubble time is given by $E = \Gamma/H^4$, where $\Gamma$ is the tunnelling 
rate per unit volume per unit time determined by the shape of the potential 
barrier between the false and true vacuum states. Provided $E$ is some way 
less than one, bubbles do not percolate and the phase transition is unable 
to complete --- the graceful exit problem of old inflation. If $E$ is much 
less than one, then any bubble collisions at all are extremely rare, 
and a single-bubble universe can survive in isolation within the surrounding 
`sea' of de Sitter space. Gott and Statler \cite{gs} found that $E < 1/400$ 
was sufficient to keep the bubbles isolated, for any $\Omega_0 \geq 0.2$.

Since $\Gamma$ is otherwise unconstrained, and since tunnelling rates 
commonly are strongly exponentially suppressed, there is no problem in us 
choosing a nucleation rate so low as to keep our scenario viable. The 
minimum value of $H$ of relevance will be that of the static point, and 
keeping $E < 1/400$ only requires 
\begin{equation}
\Gamma < \frac{4\pi^2 \, V_{{\rm fv}}^2}{225 \, {m_{{\rm Pl}}^4}} \,.
\end{equation} 

\subsection{Dynamics of the second inflationary phase}

Once $\sigma$ has tunnelled to the minimum of its potential, nucleating a 
bubble of true vacuum $V(\sigma)=0$, the $\phi$ field then rolls towards the 
minimum of its potential at $\phi^2=\nu^2$ as in single-field induced 
gravity inflation. We will take $\phi$ to be positive, and make the 
slow-roll approximations 
\begin{equation}
\dot{\phi}^2 \ll V  \quad , \quad
	\frac{\dot{\phi}^2}{\phi} \ll  3 H \dot{\phi} \quad , \quad
	\ddot{\phi} \ll 3 H \dot{\phi} \,.
\end{equation}
Dropping the curvature term, which becomes negligible once inflation has 
commenced, the equations of motion have the simple solution \cite{fu}
\begin{eqnarray}
\dot{\phi} & = & - \frac{m_{{\rm Pl}}^2}{8\pi} \, \frac{1}{1+6\xi} \,
	\sqrt{\frac{2 \lambda}{3 \xi}} \,, \\
H & = & \sqrt{\frac{ \lambda}{24 \xi}} \, \left( 1 - 
	\frac{ m_{{\rm Pl}}^2}{8 \pi \xi \phi^2} \right) \, \phi \,.
\end{eqnarray}
Meanwhile, outside the bubble $\phi$ remains at the static point with $H$ 
constant, so that the motion of the $\sigma$ and $\phi$ fields is 
synchronized and the initial hypersurface for the open universe, the bubble 
wall, remains homogeneous.

Taking $\ddot{a} > 0$, applied to the slow-roll solution, as the condition 
for inflation to occur, inflation ceases once $\phi=\phi_{{\rm end}}$, where 
\begin{equation}
\phi_{{\rm end}}^2= \frac{1 + 8 \xi + 2\sqrt{\xi(2 + 13\xi)}}{1 +6\xi}
	\; \nu^2 \,.
\end{equation}
So inflation ends slightly before $\phi$ reaches the minimum at $\phi=\nu$, 
though the difference is not important.

In the slow-roll limit the amount of inflation measured in different 
conformal frames is the same \cite{us}, so the standard expression 
for the number of $e$-foldings 
\begin{equation}
N = \int^{\phi}_{\phi_{{\rm end}}} \frac{H}{\dot{\phi}} d\phi \,,
\end{equation} 
can be applied directly. Substituting in the slow-roll solution, then to 
have 60 $e$-foldings we require
\begin{equation}
\label{efold}
\frac{\phi_{{\rm st}}^2}{m_{{\rm Pl}}^2} = \frac{\phi_{{\rm end}}^2}
	{m_{{\rm Pl}}^2} + \frac{1}{4\pi \xi} \ln 
	\frac{\phi_{{\rm st}}}{\phi_{{\rm end}}} + 
	\frac{60}{\pi(1+6 \xi)} \,.
\end{equation}
Accetta, Zoller and Turner \cite{azt} found approximate analytic solutions 
for $\phi(N)$ in each of the limits $\xi \ll 1/240$ and $\xi \gg 1/240$, 
with the additional approximation that $\phi_{{\rm end}} = \nu$. However we 
will not use this approach as values of $\xi$ between these limits are also 
interesting; instead we solve Eq.~(\ref{efold}) numerically by iteration.

The false vacuum energy of the $\sigma$ field can be calculated from the 
condition that there are 60 $e$-foldings of inflation in the second phase, 
giving
\begin{equation}
V_{{\rm fv}} = \frac{\lambda m_{{\rm Pl}}^4}{64 \pi \xi} \left(
	\frac{\phi_{{\rm st}}^2}{m_{{\rm Pl}}^2} - \frac{1}{8 \pi \xi}
	\right) \,.
\end{equation}

\section{Density Perturbations}

There are three recognized types of density perturbations in open inflation 
models \cite{yst}, sub-curvature modes, super-curvature modes and modes 
associated with bubble wall fluctuations.

\subsection{Sub-curvature modes}

Except on the largest scales, perturbations can be considered to have formed 
during the second phase of inflation. The standard calculation requires 
Einstein gravity and a canonically normalized scalar field, which is brought 
about by the conformal transformation Eq.~(\ref{conftran}) and field 
redefinition Eq.~(\ref{fielddef}). Then the standard formula for 
the density perturbations can be applied, namely \cite{LL}
\begin{equation}
\delta_{{\rm H}} = \frac{1}{5 \pi} \frac{\tilde{H}^2}
	{|d \tilde{\phi}/d\tilde{t}|} \,,
\end{equation}
where tildes refer to quantities in the Einstein frame. The quantities on 
the right-hand side should be evaluated when the relevant scales crossed 
outside the Hubble radius during inflation. The Einstein frame 
quantities can be related to those in the original frame without explicitly 
performing the conformal transformation, by using
\begin{eqnarray}
\tilde{H} & = & \frac{1}{C} H + \frac{1}{C^2} \frac{dC}{dt} \,, \\
\frac{d\tilde{\phi}}{d\tilde{t}} & = & 	\frac{m_{{\rm Pl}}^2}{8\pi \phi^2}
	\; \frac{\sqrt{1+6\xi}}{\xi} \; \frac{d\phi}{dt} \,.
\end{eqnarray}
Employing the slow-roll approximation, which allows us to drop the second 
term in the expression for $\tilde{H}$, we obtain
\begin{equation}
\label{perts}
\delta_{{\rm H}} = \sqrt{\frac{\lambda}{150}} \,
	\sqrt{\frac{1+6\xi}{\xi}} \, \frac{\phi^2}{m_{{\rm Pl}}^2} \, 
	\left( 1 - \frac{m_{{\rm Pl}}^2}{8 \pi \xi \phi^2} \right)^2 \,.
\end{equation}

The microwave anisotropies seen by COBE \cite{COBE} correspond to 
perturbations generated about 60 $e$-foldings from the end of inflation, 
obtained by substituting $\phi \simeq \phi_{{\rm st}}$ in Eq.~(\ref{perts}). 
In a spatially flat universe, the observed amplitude is reproduced if 
$\delta_{{\rm H}} \simeq 2 \times 10^{-5}$ \cite{BW}. The required value 
actually changes somewhat if $\Omega_0$ is significantly less than one 
\cite{BW}, but at the level of accuracy we are working we can ignore this 
correction. Obtaining the correct density perturbations determines, for a 
given $\xi$, the value of $\lambda$ required.  

\subsection{Super-curvature and bubble wall modes}

In models where the mass in the false vacuum obeys $m^2 < 2H^2$, one 
typically expects a discrete super-curvature mode \cite{sty,yst}. The 
potential for $\sigma$ is unconstrained in this respect, and so may or may 
not support such a mode. Much more important is the dilaton; considering the 
Einstein frame quantities, and denoting the ratio by $\mu$, we have
\begin{equation}
\mu^2 \equiv \left. \frac{m^2_{\tilde{\phi}}}{\tilde{H}^2}
	\right|_{\phi = \phi_{{\rm st}}} = \frac{3\xi}{1+6\xi} \,
	\frac{\lambda \nu^4}{V_{{\rm fv}}} \,,
\end{equation}
which is typically small and so a super-curvature mode from the $\phi$ 
field is expected.\footnote{We thank Juan Garc\'{\i}a-Bellido for pointing 
this out to us.}

For computing this mode, our situation is extremely similar to the 
`supernatural' model of Linde and Mezhlumian \cite{lin}; the fields are 
initially both trapped, and then one becomes free to roll after tunnelling. 
The full spectrum of modes in the supernatural model was computed by 
Yamamoto et al.~\cite{yst}, in the limit where the expansion rate before and 
after tunnelling is taken to be the same. As we shall see, we are typically 
in this limit. They find that while there is a contribution from the 
super-curvature mode, it is never very large regardless of $\mu$ and so does 
not threaten the viability of that model. A detailed calculation for our 
model is outside the scope of this paper, but we expect that the result is 
extremely similar, so long as the assumption that the expansion rate before 
and after tunnelling is equal holds.

Sasaki and Tanaka \cite{st} have made detailed calculations of the situation 
where the expansion rate changes significantly during tunnelling, with 
particular reference to the model of Eq.~(\ref{linpot}). They found that the 
super-curvature fluctuations are enhanced, relative to the sub-curvature 
ones, by a factor of order $(H_{{\rm false}}/H_{{\rm true}})^2$, where these 
are the expansion rates before and after tunnelling. Such an enhancement 
could distort the cosmic microwave background (CMB) spectrum from that 
observed, since the super-curvature modes only contribute significantly to 
anisotropies with multipole number $l \lesssim 10$ \cite{GZpaper}. 

In our model this factor is given by 
\begin{equation}
\left( \frac{H_{{\rm false}}}{H_{{\rm true}}} \right)^2 = 
	\frac{\lambda m_{{\rm Pl}}^2}{ 8 \xi V_{{\rm fv}}} \;
	\frac{(\phi^2 - \nu^2)^2}{\phi^2} \,.
\end{equation}
Evaluating this expression immediately after tunnelling gives
\begin{equation}
\left( \frac{H_{{\rm false}}}{H_{{\rm true}}} \right)^2  = 1 +
	\frac{\lambda \nu^4}{8 V_{{\rm fv}}} \,,
\end{equation}
which leads to a limit on the ratio of the false vacuum energies of the 
$\phi$ and $\sigma$ fields.

Compared to the model of Eq.~(\ref{linpot}), we have the considerable 
advantage that the fields outside the bubble are already static, rather than 
having to be damped by a large difference in the expansion rate. As long as 
$\xi$ is not too small, the ratio of energy densities is never big and so  
enhancement of the super-curvature modes will not be a problem for us. This 
is also the case with other models proposed in Ref.~\cite{lin}.

The final type of density perturbations are those associated with 
fluctuations in the bubble wall \cite{bubble,yst}. These will be no 
different from those in other types of open inflation model, where they have 
not proven constraining.

\section{Parameters}
 
For any particular value of $\Omega_0$, we can freely choose $\xi$ but 
then the rest of the parameters of the model are uniquely determined: $\nu$ 
from reproducing the present-day Planck mass, $V_{{\rm fv}}$ (in terms 
of $\lambda$) from the required position of $\phi_{{\rm st}}$, then 
$\lambda$ from the magnitude of the density perturbations. Finally, an upper 
bound is placed on $\Gamma$ from the requirement that the true-vacuum 
bubbles do not percolate.

In the limit $\xi \ll 1/240$, we have $\phi_{{\rm st}} \simeq \nu$ and can 
solve Eq.~(\ref{efold}) by small parameter expansion. The parameters have 
approximate analytic forms
\begin{eqnarray}
\lambda & \approx & 4.1 \times 10^{-11} \xi \,, \\ 
V_{{\rm fv}} & \approx & 2.5 \times 10^{-13} \xi^{-1/2} \, 
	m_{{\rm Pl}}^4\,, \\
\Gamma & < & 1.1 \times 10^{-26} \xi^{-1} m_{{\rm Pl}}^4 \,, \\
\left( \frac{H_{{\rm false}}}{H_{{\rm true}}} \right)^2 & \approx & 1 +
	0.03 \, \xi^{-1/2} \,, \\
\mu^2 & = & 0.8 \, \xi^{1/2} \,.
\end{eqnarray}

\begin{figure}[t]
\centering 
\leavevmode\epsfysize=16cm \epsfbox{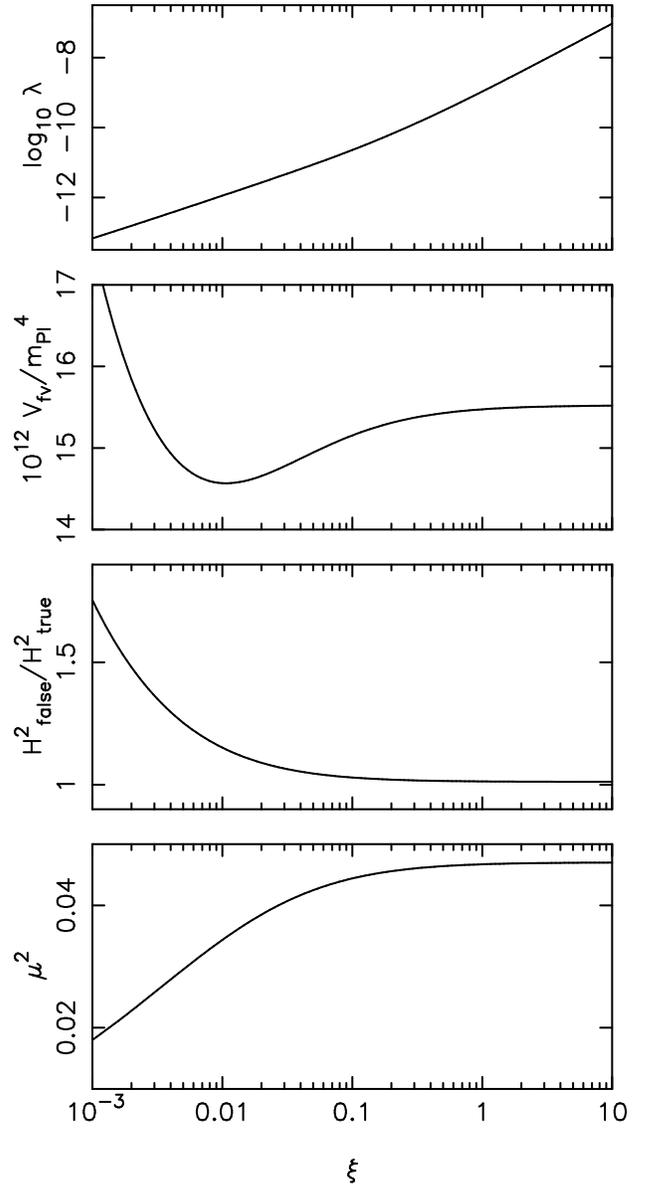}\\ 
\caption[fig3]{\label{fig3} The principal parameters as functions of $\xi$, 
for COBE-normalized models and assuming 60 $e$-foldings in the second 
phase. For $\xi \ll 1/240$, the analytic forms given in the text apply.} 
\end{figure} 

Away from that limit, we must solve Eq.~(\ref{efold}) numerically. For a 
sample value of $\xi = 10^{-2}$ we find
\begin{eqnarray}
\lambda & = & 1.1 \times 10^{-12} \,, \\
V_{{\rm fv}} & = & 1.4 \times 10^{-11} \, m_{{\rm Pl}}^4 \,, \\
\Gamma & < & 3.6 \times 10^{-23} \, m_{{\rm Pl}}^4 \,, \\ 
\left( \frac{H_{{\rm false}}}{H_{{\rm true}}} \right)^2 & = & 1.15 \,, \\
\mu^2 & = & 0.03 \,.
\end{eqnarray}
In Fig.~3, we plot the values of the different parameters as functions of 
$\xi$, under the assumption of $60$ $e$-foldings in the second phase. The 
COBE normalization fixes the other quantities as functions of $\xi$ alone 
once the number of $e$-foldings has been fixed. As we stated earlier, the 
precise value of $\Omega_0$ this corresponds to depends on the reheat energy 
and other related uncertainties, but clearly the numbers will not change 
much if the number of $e$-foldings is varied somewhat to cover the entire 
interesting range of $\Omega_0$ values.

Notice that these are all very reasonable values. The coupling $\xi$ to 
gravity can easily be of order one or less, and the false vacuum energy is 
around the standard unification energy. Notice by how little $V_{{\rm fv}}$ 
changes as $\xi$ is varied. The self-coupling $\lambda$ does need to be 
small in order to give the correct magnitude of density perturbations, but 
this is a familiar requirement from a wide range of inflation models.

\section{Summary}

We have constructed a new model of open inflation, based on the induced 
gravity theory, which capitalizes on the existence of a static, stable de 
Sitter solution in the presence of a false vacuum. We have shown that this 
model can easily satisfy all known constraints, both from inflationary 
dynamics and from the production of density perturbations. The value of 
$\Omega_0$ is uniquely determined from the model parameters, though 
uncertainties related to reheating prevent an accurate calculation at 
present. It would be extremely interesting to see the results of a full 
calculation of all types of density perturbations in this model, along the 
lines of Ref.~\cite{yst}.

\section*{Acknowledgments}

A.~M.~G.~is supported by PPARC and A.~R.~L.~by the Royal Society. We
are indebted to Juan Garc\'{\i}a-Bellido for a series of discussions on 
super-curvature modes, and also thank Jim Lidsey, Andrei Linde and David 
Wands for their comments. We acknowledge use of the Starlink computer system 
at the University of Sussex. 


\begin{references}
\bibitem{gott} J. R. Gott, Nature {\bf 295}, 304 (1982).
\bibitem{gs} J. R. Gott	and T. S. Statler, Phys. Lett. {\bf 136B}, 
	157 (1984).
\bibitem{gott2} J. R. Gott in {\em Inner Space, Outer Space}, eds. E. W.
	Kolb, M. S. Turner, D. Lindley, K. Olive and D. Seckel, 
	University of Chicago Press (1986).
\bibitem{styy} M. Sasaki, T. Tanaka, K. Yamamoto and J. Yokoyama, 
	Phys. Lett. B {\bf 317}, 510 (1993).
\bibitem{bt} M. Bucher, A. S. Goldhaber and N. Turok, Phys. Rev. D 
	{\bf 52}, 3314; M. Bucher and N. Turok, Phys. Rev. D {\bf 52},
	5538 (1995).
\bibitem{lin} A. D. Linde, Phys. Lett. B {\bf 351}, 99 (1995); A. D.
	Linde and A. Mezhlumian, Phys. Rev. D {\bf 52}, 6789 (1995).
\bibitem{CdL} S. Coleman and F. de Luccia, Phys. Rev. D {\bf 21}, 3305
	(1980).
\bibitem{st} M. Sasaki and T. Tanaka, e-print archive:
	astro-ph/9605104 (1996).
\bibitem{z} A. Zee, Phys. Rev. Lett. {\bf 42}, 417 (1979).
\bibitem{bd} C. Brans and R. H. Dicke, Phys. Rev. {\bf 24}, 925 (1961).
\bibitem{will} C. M. Will, {\em Theory and Experiment in Gravitational
	Physics}, Cambridge University Press (Cambridge, 1993).
\bibitem{COBE} C. L. Bennett et al., Astrophys. J. Lett. {\bf 464}, L1
	(1996).
\bibitem{azt} F. S. Accetta, D. J. Zoller and M. S. Turner,
	Phys. Rev. D {\bf 31}, 3046 (1985).
\bibitem{Kaiser} D. I. Kaiser, Phys. Rev. D {\bf 49}, 6347 (1994).
\bibitem{fu} R. Fakir and W. G. Unruh, Phys. Rev. D {\bf 41}, 1792
	(1990).
\bibitem{at} F. S. Accetta and J. J. Trester, Phys. Rev. D {\bf 39}, 2854
	(1989).
\bibitem{us} A. M. Green and A. R. Liddle, to appear, Phys. Rev. D,
	astro-ph/9604001 (1996).
\bibitem{yst} K. Yamamoto, M. Sasaki and T. Tanaka, e-print archive:
	astro-ph/9605103 (1996).
\bibitem{LL} A. R. Liddle and D. H. Lyth, Phys. Rep. {\bf 231}, 1 (1993).
\bibitem{BW} E. F. Bunn and M. White, e-print archive: astro-ph/9607060
	(1996).
\bibitem{sty} M. Sasaki, T. Tanaka and K. Yamamoto, Phys. Rev. 
	D {\bf 51}, 2979 (1995).
\bibitem{GZpaper} J. Garc\'{\i}a-Bellido, A. R. Liddle, D. H. Lyth and
	D. Wands, Phys. Rev. D {\bf 52}, 6750 (1995).
\bibitem{bubble} J. Garc\'{\i}a-Bellido, to appear, Phys. Rev. D,
	astro-ph/9510029 (1995); J. Garriga, to appear, Phys. Rev. D,
	gr-qc/9602025 (1996).
\end{references}
\end{document}